\documentclass[aps,pra,showpacs,superscriptaddress]{revtex4}
\usepackage{amsfonts}
\usepackage{amsmath}
\usepackage{amssymb}
\usepackage{graphicx}

\begin{document}

\title{Vortex states and spin textures of rotating spin-orbit-coupled
Bose-Einstein condensates in a toroidal trap}
\author{Huan Wang}
\affiliation{College of Science, Yanshan University, Qinhuangdao 066004, China}
\author{Linghua Wen}
\email{linghuawen@ysu.edu.cn}
\affiliation{College of Science, Yanshan University, Qinhuangdao 066004, China}
\author{Hui Yang}
\affiliation{College of Science, Yanshan University, Qinhuangdao 066004, China}
\author{Chunxiao Shi}
\affiliation{College of Science, Yanshan University, Qinhuangdao 066004, China}
\author{Jinghong Li}
\affiliation{College of Environment and Chemical Engineering,Yanshan University,
Qinhuangdao 066004, China}
\date{\today }

\begin{abstract}
We consider the ground-state properties of Rashba spin-orbit-coupled
pseudo-spin-1/2 Bose-Einstein condensates (BECs) in a rotating
two-dimensional (2D) toroidal trap. In the absence of spin-orbit coupling
(SOC), the increasing rotation frequency enhances the creation of giant
vortices for the initially miscible BECs, while it can lead to the formation
of semiring density patterns with irregular hidden vortex structures for the
initially immiscible BECs. Without rotation, strong 2D isotropic SOC yields
a heliciform-stripe phase for the initially immiscible BECs. Combined
effects of rotation, SOC, and interatomic interactions on the vortex
structures and typical spin textures of the ground state of the system are
discussed systematically. In particular, for fixed rotation frequency above
the critical value, the increasing isotropic SOC favors a visible vortex
ring in each component which is accompanied by a hidden giant vortex plus a
(several) hidden vortex ring(s) in the central region. In the case of 1D
anisotropic SOC, large SOC strength results in the generation of hidden
linear vortex string and the transition from initial phase separation (phase
mixing) to phase mixing (phase separation). Furthermore, the peculiar spin
textures including skyrmion lattice, skyrmion pair and skyrmion string are
revealed in this system.
\end{abstract}

\maketitle

\section{Introduction}

Spin-orbit coupling (SOC) of a quantum particle plays a key role in many
physical phenomena, including fine structures of atomic spectra, spin Hall
effect \cite{Xiao}, topological insulators \cite{Qi} and topological
superconductors \cite{Read}. In general, the SOC strengths in solid
materials are fixed, and the observation of SOC effects is hindered by
unavoidable impurities and disorder. In contrast, the spin-orbit coupled
ultracold atomic gases demonstrated by the milestone experiments \cite%
{Lin,PWang,Cheuk,JYZhang} offer an extremely clean platform with full
controllability to explore novel macroscopic quantum phenomena and quantum
topological states \cite{Dalibard,Zhai}, especially for bosons \cite%
{Ho,Sinha,Hu,Li,Kawakami,YZhang,Ruokokoski,Zhu}. Recently, the
two-dimensional (2D) SOCs for boson gases and for fermion gases have been
realized, respectively \cite{LHuang,ZWu}, which is crucial in studying the
exotic properties of high-dimensional topological matter. Of particular
interest are the topological excitations in rotating Bose-Einstein
condensates (BECs) with SOC. Relevant studies show that the spin-$1/2$ BECs
with SOC in a rotating harmonic trap support the formation of half-quantum
vortex, giant vortex, skyrmion, and multidomain pattern \cite%
{Xu,Radic,Zhou,Ramachandhran,Liu,Aftalion,ZFXu,Fetter}.

However, there are few studies so far on the ground-state properties of
rotating spin-orbit-coupled BECs in a toroidal trap. In fact, various
trapping potentials may significantly affect the ground states and dynamic
properties of the BECs \cite{Ueda,Wu,Wen1,Kasamatsu,Wen2}, which leads to a
rich physics. In this work, we investigate the vortex states and spin
textures of rotating BECs with Rashba SOC in a toroidal trap. As a
non-trivial geometry, toroidal trap can be easily created in current
experiments \cite{Ryu,Beattie,Corman,Eckel,Wood}, and it provides us a
unique platform to investigate the exotic properties of quantum superfluids.
Therefore our work is experimentally feasible and allows to be tested in
future experiments. We show that increasing 2D isotropic SOC can enhance the
formation of visible vortex ring, hidden giant vortex (or hidden singly
quantized vortices), and spin domain walls for large rotation frequency
above the critical value. In the nonrotating limit, strong 2D isotropic SOC
generates a heliciform-stripe phase for the initially separated condensates
in a toroidal trap. In addition, large strength of the 1D SOC leads to the
creation of hidden linear vortex string and the transition from initial
phase separation (or phase mixing) to phase mixing (or phase separation).
Moreover, the combined effects of the rotation, SOC, and interparticle
interactions on the ground-state properties of the system are discussed.

The paper is organized as follows. In section $2$, the model Hamiltonian is
introduced and the coupled dynamic equations are given. The topological
structures and typical spin textures of the ground state of the system are
described and analyzed in section $3$. Conclusions are outlined in section $%
4 $.

\section{Theoretical model}

We consider a quasi-two-dimensional $(x,y)$ rotating spin-orbit-coupled BEC
in a toroidal trap. The model Hamiltonian for the interacting pseudo-spin-$%
1/2$ BEC is given by
\begin{equation}
\hat{H}=\int dxdy\hat{\psi}^{\dag }\left[ -\frac{\hbar ^{2}\bigtriangledown
^{2}}{2m}+V(r)+g_{1}\hat{n}_{1}^{2}+g_{2}\hat{n}_{2}^{2}+2g_{12}\hat{n}_{1}%
\hat{n}_{2}-\Omega L_{z}+v_{SO}\right] \hat{\psi},  \label{Hamiltonian}
\end{equation}%
where $\hat{\psi}=[\hat{\psi}_{1}(r),\hat{\psi}_{2}(r)]^{T}$ denotes
collectively the spinor Bose field operators with $1$ and $2$ corresponding
to spin-up and spin-down, respectively. $\hat{n}_{1}=\hat{\psi}_{1}^{\dag }%
\hat{\psi}_{1}$ and $\hat{n}_{2}=\hat{\psi}_{2}^{\dag }\hat{\psi}_{2}$ are
the density operators of the particle numbers for spin-up and spin-down
atoms, respectively. $g_{j}=4\pi a_{j}\hbar ^{2}/m$ $($ $j=1,2)$ and $%
g_{12}=2\pi a_{12}\hbar ^{2}/m$ represent the intra- and intercomponent
coupling strengths, where $m$ is the atomic mass, $a_{j}$ $($ $j=1,2)$ and $%
a_{12}$ are the $s$-wave scattering lengths between intra- and
intercomponent atoms. $\Omega $ is the rotation angular velocity along the $%
z $ direction, and $\hat{L}_{z}=-i\hbar (x\partial _{y}-y\partial _{x})$ is
the $z$ component of the angular-momentum operator. The Rashba SOC term is
given by $v_{SO}=-i\hbar (\lambda _{y}\hat{\sigma}_{x}\partial _{y}-\lambda
_{x}\hat{\sigma}_{y}\partial _{x})$ with $\hat{\sigma}_{x,y}$ being Pauli
matrices and $\lambda _{x}$ and $\lambda _{y}$ being the SOC strengths in
the $x$ and $y$ directions. The external toroidal potential is expressed as
\cite{Cozzini}
\begin{equation}
V(r)=\frac{1}{2}m\omega _{\perp }^{2}\left[ V_{0}(\frac{r^{2}}{a_{0}}%
-a_{0}r_{0})^{2}\right] =\frac{1}{2}\hbar \omega _{\perp }\left[ V_{0}(\frac{%
r^{2}}{a_{0}^{2}}-r_{0})^{2}\right] ,  \label{Toroidal potential}
\end{equation}%
where $\omega _{\perp }$ is the radial oscillation frequency, $a_{0}=\sqrt{%
\hbar /m\omega _{\perp }}$, $r=\sqrt{x^{2}+y^{2}}$. $V_{0}$ and $r_{0}$ are
dimensionless constants which characterize the width and the central height
of the toroidal potential. $(\pm a_{0}\sqrt{r_{0}},0)$ and $(0,\hbar \omega
_{\perp }V_{0}r_{0}^{2}/2)$ represent the lowest point and the highest one
of the potential well, respectively. The normalization condition of the
system reads
\begin{equation}
\int \left[ |\psi _{1}|^{2}+|\psi _{2}|^{2}\right] dxdy=N,
\label{Normalization}
\end{equation}%
where $\psi _{j}$ $($ $j=1,2)$ is the wave function of component $j$, and $N$
is the number of atoms. By introducing the notations $\widetilde{t}=\omega
_{\perp }t$, $\widetilde{r}=r/a_{0}$, $\widetilde{V}=V/\hbar \omega _{\perp
} $, $\widetilde{\Omega }=\Omega /\omega _{\perp }$, $\widetilde{L}%
_{z}=L_{z}/\hbar $, and $\widetilde{\psi }_{j}=\psi _{j}a_{0}/\sqrt{N}($ $%
j=1,2)$, we obtain the dimensionless coupled Gross-Pitaevskii (GP) equations
for the dynamics of the system in terms of the variational principle,
\begin{equation}
i\partial _{t}\psi _{1}=(-\frac{1}{2}\bigtriangledown ^{2}+V+\beta
_{11}|\psi _{1}|^{2}+\beta _{12}|\psi _{2}|^{2})\psi _{1}-\Omega L_{z}\psi
_{1}+(\lambda _{x}\partial _{x}-i\lambda _{y}\partial _{y})\psi _{2},
\label{Component1}
\end{equation}%
\begin{equation}
i\partial _{t}\psi _{2}=(-\frac{1}{2}\bigtriangledown ^{2}+V+\beta
_{12}|\psi _{1}|^{2}+\beta _{22}|\psi _{2}|^{2})\psi _{2}-\Omega L_{z}\psi
_{2}-(\lambda _{x}\partial _{x}+i\lambda _{y}\partial _{y})\psi _{1},
\label{Component2}
\end{equation}%
where the tilde is omitted for simplicity. $\beta _{jj}$ $($ $j=1,2)$ and $%
\beta _{12}=\beta _{21}$ are the dimensionless intra- and intercomponent
interaction strengths.

In the frame of nonlinear Sigma model \cite{Aftalion,Mizushima,Kasamatsu2},
we can introduce a normalized complex-valued spinor $\mathbf{\chi }=[\chi
_{1},\chi _{2}]^{T}$ with $|\chi _{1}|^{2}+|\chi _{2}|^{2}=1$. The total
density of the system is given by $\rho =|\psi _{1}|^{2}+|\psi _{2}|^{2}$,
and the corresponding two-component wave functions are $\psi _{1}=\sqrt{\rho
}\chi _{1}$ and $\psi _{2}=\sqrt{\rho }\chi _{2}$, respectively. The spin
density is defined as $\mathbf{S}=\overline{\mathbf{\chi }}\mathbf{\sigma
\chi }$, where $\mathbf{\sigma }=(\sigma _{x},\sigma _{y},\sigma _{z})$ are
the Pauli matrices, and the components of $\mathbf{S}$ are expressed as
\begin{eqnarray}
S_{x} &=&\chi _{1}^{\ast }\chi _{2}+\chi _{2}^{\ast }\chi _{1},  \label{Sx}
\\
S_{y} &=&i(\chi _{2}^{\ast }\chi _{1}-\chi _{1}^{\ast }\chi _{2}),
\label{Sy} \\
S_{z} &=&|\chi _{1}|^{2}-|\chi _{2}|^{2},  \label{Sz}
\end{eqnarray}%
with $|\mathbf{S}|^{2}=S_{x}^{2}+S_{y}^{2}+S_{z}^{2}=1$. In order to
describe the spacial distribution of the topological structure, we introduce
a topological charge density%
\begin{equation}
q(r)=\frac{1}{4\pi }\mathbf{S\bullet }\left( \frac{\partial \mathbf{S}}{%
\partial x}\times \frac{\partial \mathbf{S}}{\partial y}\right) ,
\label{TopologicalChargeDensity}
\end{equation}%
then we can obtain the topological charge $Q$ by calculating the whole space
integral of $q(r)$,%
\begin{equation}
Q=\int q(r)dxdy.  \label{TopologicalCharge}
\end{equation}

\section{Results and discussion}

In what follows, we numerically solve the 2D coupled GP equations (\ref%
{Component1}) and (\ref{Component2}). We obtain the minimizing energy state
(i.e., the ground state) of the system by using the imaginary-time
propagation method \cite{YZhang} based on the Peaceman-Rachford method \cite%
{Peaceman,Wen3}. Recently, a phase diagram for a nonrotating
spin-orbit-coupled BEC in a harmonic trap has been discussed in the
literature \cite{Aftalion}. In the present work, we investigate
systematically the combined effects of rotation, SOC and interatomic
interactions on the ground state of the BECs in a toroidal trap. In our
simulation, the typical parameters of the toroidal potential are chosen as $%
V_{0}=0.5$ and $r_{0}=3$, the intraspecies interactions are fixed as $\beta
_{11}=\beta _{22}=100$. It is shown that system can exhibit rich and exotic
topological structures and spin textures due to the topology difference
between the toroidal and the harmonic traps.

\subsection{Rotation effect}

Firstly, we consider the effect of rotation on the ground state of a
toroidal interacting spin-1/2 BEC in the absence of SOC. Figure 1 shows the
density distributions [(a), (b) and (d)] and phase distributions [(c) and
(e)] of the ground states of the system, where $\beta _{11}=\beta _{22}=100$%
, $\beta _{12}=50$ (the top two rows), and $\beta _{12}=150$ (the bottom two
rows). The rotation frequencies are $\Omega =0$ [column (a)], $\Omega =0.8$
[the top two rows of (b)-(c)], $\Omega =1.2$ [the bottom two rows of
(b)-(c)], and $\Omega =2$ ((d)-(e)), respectively. Here the odd and even
rows denote component $1$ and component $2$, respectively.

\begin{figure}[tbh]
\centerline{\includegraphics*[width=10cm]{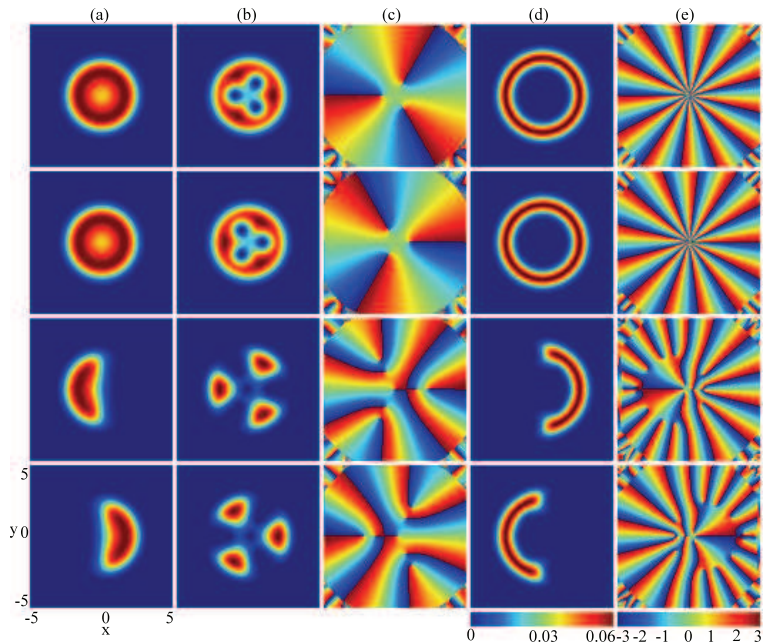}}
\caption{(Color online) Density distributions [(a), (b), and (d)] and phase
distributions [(c) and (e)] for the ground states of rotating toroidal spin-$%
1/2$ BECs in the absence of SOC. Here the odd rows correspond to component $%
1 $, while the even ones correspond to component $2$. The interaction
parameters are $\protect\beta _{11}=\protect\beta _{22}=100$, $\protect\beta %
_{12}=\protect\beta _{21}=50$ (the top two rows) and $\protect\beta _{12}=%
\protect\beta _{21}=150$ (the bottom two rows), which correspond to phase
mixing and phase separation of a nonrotating system, respectively. The
rotation frequency are $\Omega =0$ [column (a)], $\Omega =0.8$ [the top two
rows of (b)-(c)], $\Omega =1.2$ [the bottom two rows of (b)-(c)], and $%
\Omega =2$ [(d)-(e)]. The horizontal and vertical coordinates $x$ and $y$
are in units of $a_{0}$.}
\label{Figure1}
\end{figure}

As shown in figure 1(a), the two components of the nonrotating BEC display
typical phase mixing (top two rows) and phase separation (bottom two rows),
due to the interatomic interactions satisfying $\beta _{11}\beta _{22}>\beta
_{12}^{2}$ and $\beta _{11}\beta _{22}<\beta _{12}^{2}$, respectively. For
the convenience of discussion, in the following texts we call the two phases
briefly initial phase mixing and initial phase separation, respectively. In
the case of initial phase mixing, when $\Omega =0.8$ the rotation driving
inputs sufficient angular momentum such that three visible vortices \cite%
{Wen1,Wen3} occur in each component and constitute a triangular vortex
lattice (Abrikosov lattice), where the two visible vortex lattices are
staggered each other [see the top two rows in (b)-(c)]. When $\Omega =2$ a
giant vortex (a multiply-quantized vortex) is generated in each component
[see the top two rows of (d)-(e)], which is inaccessible in a harmonically
trapped BEC because for $\Omega \rightarrow 1$ the resulting centrifugal
effect would cancel the radial confinement and the Thomas-Fermi (TF) radius
of the BEC would diverge. In the case of initial phase separation, when $%
\Omega =1.2$ the component densities become separated petals, where the
hidden vortices \cite{Wen1,Wen3,Mithun,Wen4} form a regular vortex cluster.
With the further increase of rotation frequency, e.g., $\Omega =2$, the
component densities form two butt-joint semi-rings and the hidden vortices
display complex topological structures, which can be seen in the bottom two
rows of (d)-(e).

\subsection{The effect of SOC}

Secondly, we consider the effect of SOC on the ground state of the
nonrotating spin-1/2 BEC in a toroidal potential. The main results are
illustrated in figure 2. For 2D weak SOC, e.g., $\lambda _{x}=\lambda _{y}=1$%
, the ground-state structure of the system with an initial phase mixing is
similar to that with an initial phase separation except for the difference
between the density mixing and the density separation, where few ghost
vortices are generated in each component [see the left four columns in
(a)-(b)]. However, for 2D strong SOC, e.g., $\lambda _{x}=\lambda _{y}=10$,
there exists a remarkable difference between the ground-state structure in
the case of initial phase mixing and that in the case of initial phase
separation [see the left four columns in (c)-(d)]. For the former case, a
visible vortex string is formed along the $y=0$ axis in each component, and
the two vortex strings are separated spatially due to the strong SOC. In
contrast, for the latter case, the two components display spatially
separated and heliciform stripe structures. When there is only 1D SOC in the
BEC, e.g., $\lambda _{x}=10$ and $\lambda _{y}=0$, the system exhibits
either a\ fully mixed phase with a fringe-shape phase distribution for weak
interspecies repulsion or a separated stripe phase along the $x$ direction
for strong interspecies repulsion, which are referred as Thomas-Fermi (TF)
phase (plane wave phase) and stripe phase \cite{YZhang}, respectively. As
for the other values of $\lambda _{x}$ and $\lambda _{y}=0$, our simulation
shows that the density and phase patterns are similar to those in the right
two columns of figure 2. If the 1D SOC is along the $y$ direction, e.g., $%
\lambda _{x}=0$ and $\lambda _{y}=10$, the density and phase distributions
will correspondingly rotate $90$ degrees due to the rotation of the SOC.

\begin{figure}[tbh]
\centerline{\includegraphics*[width=13cm]{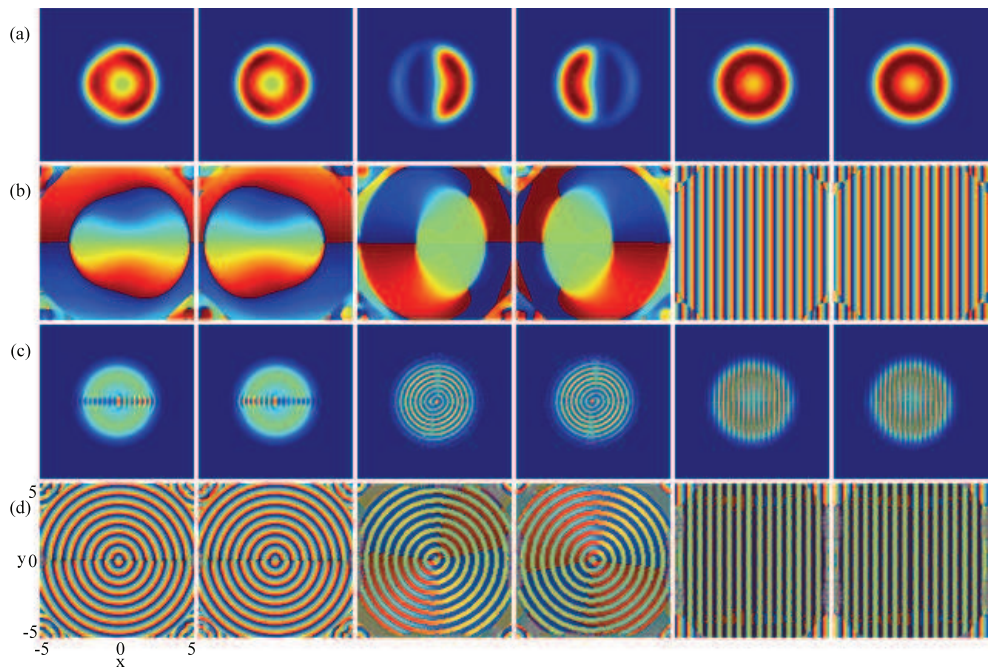}}
\caption{(Color online) Density distributions [(a) and (c)] and phase
distributions [(b) and (d)] for the ground states of nonrotating SOC spin-$%
1/2$ BECs in a toroidal trap. The odd and even columns correspond to
component $1$ and component $2$, respectively. The left two columns: $%
\protect\beta _{12}=50$, (a)-(b) $\protect\lambda _{x}=\protect\lambda %
_{y}=1 $, and (c)-(d) $\protect\lambda _{x}=\protect\lambda _{y}=10$. The
middle two columns: $\protect\beta _{12}=150$, (a)-(b) $\protect\lambda _{x}=%
\protect\lambda _{y}=1$, and (c)-(d) $\protect\lambda _{x}=\protect\lambda %
_{y}=10$. The right two columns: $\protect\lambda _{x}=10$, $\protect\lambda %
_{y}=0$, (a)-(b) $\protect\beta _{12}=50$, and (c)-(d) $\protect\beta %
_{12}=150$. The horizontal and vertical coordinates $x$ and $y$ are in units
of $a_{0}$.}
\label{Figure2}
\end{figure}

\subsection{The combined effects of rotation, SOC and interatomic
interactions}

\subsubsection{Fixed rotation frequency}

Next, we investigate the combined effects of SOC, rotation and interatomic
interactions on the ground state of the system. Figure 3 shows the density
distributions and phase distributions for the ground states of rotating
toroidal spin-$1/2$ BECs, where $\Omega =0.9$. The strengths of the 2D SOC
for the initial phase mixing with $\beta _{12}=50$ in rows (a)-(b) are $%
\lambda _{x}=\lambda _{y}=1$ and $\lambda _{x}=\lambda _{y}=10$, and those
for the initial phase separation with $\beta _{12}=150$ in rows (c)-(e) are $%
\lambda _{x}=\lambda _{y}=0.2$, $\lambda _{x}=\lambda _{y}=1$ and $\lambda
_{x}=\lambda _{y}=8$, respectively. The columns (from left to right)
represent $|\psi _{1}|^{2}$, $|\psi _{2}|^{2}$ , arg$\psi _{1}$, arg$\psi
_{2}$, $|\psi _{1}|^{2}+|\psi _{2}|^{2}$, and $|\psi _{1}|^{2}-|\psi
_{2}|^{2}$, respectively.

\begin{figure}[tbh]
\centerline{\includegraphics*[width=13cm]{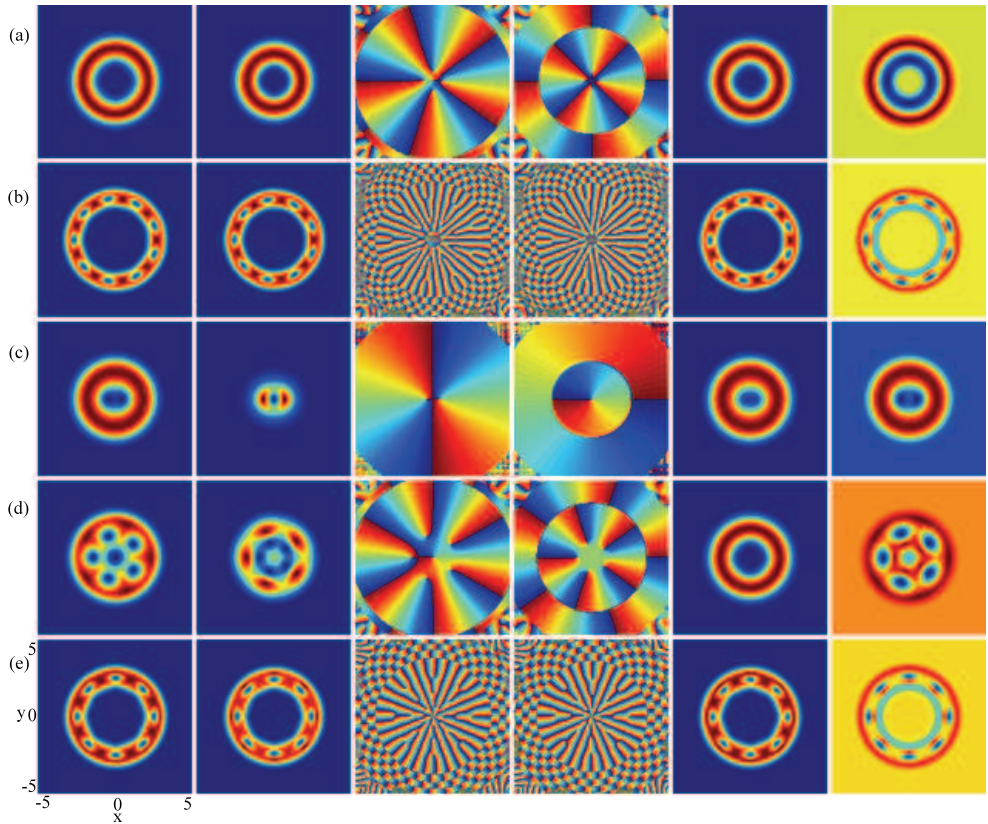}}
\caption{(Color online) Ground states of rotating interacting spin-$1/2$
BECs with SOC in a toroidal trap, where $\Omega =0.9$. (a) $\protect\beta %
_{12}=50$, $\protect\lambda _{x}=\protect\lambda _{y}=1$, (b) $\protect\beta %
_{12}=50$, $\protect\lambda _{x}=\protect\lambda _{y}=10$, (c) $\protect%
\beta _{12}=150$, $\protect\lambda _{x}=\protect\lambda _{y}=0.2$, (d) $%
\protect\beta _{12}=150$, $\protect\lambda _{x}=\protect\lambda _{y}=1$, and
(e) $\protect\beta _{12}=150$, $\protect\lambda _{x}=\protect\lambda _{y}=8$%
. The columns (from left to right) represent $|\protect\psi _{1}|^{2}$, $|%
\protect\psi _{2}|^{2}$ , arg$\protect\psi _{1}$, arg$\protect\psi _{2}$, $|%
\protect\psi _{1}|^{2}+|\protect\psi _{2}|^{2}$, and $|\protect\psi %
_{1}|^{2}-|\protect\psi _{2}|^{2}$, respectively. The horizontal and
vertical coordinates $x$ and $y$ are in units of $a_{0}$.}
\label{Figure3}
\end{figure}

As shown in figures 3(a) and 3(c), the density distributions and phase
distributions for relatively weak SOC\ and large rotation frequency display
remarkable difference between the case of initial phase mixing and that of
initial phase separation. In figure 3(a), the two large density holes in the
two components are not the normal giant vortices observed in rotating
harmonically trapped BECs with SOC \cite{Xu,Zhou} but a hidden triangular
vortex lattice and a vortex ring, respectively. While in figure 3(c) the two
component densities exhibit obvious phase separation, where the vortices in
the two components develop into an interlaced vortex array. For large SOC,
we observed that there was an enhanced overlap of the density distributions
and the phase distributions between the two components. This feature exists
not only for the case of $\beta _{12}<\beta _{11}$ but also for the case of $%
\beta _{12}>\beta _{11}$ [see figures 3(b), 3(d) and (3e)]. Here the visible
vortices form ringlike structures. The region of the large density hole is
occupied by a central hidden giant vortex and a (several) hidden vortex
ring(s) [see figures 3(b) and 3(e)]. The ringlike visible vortex
configuration has also been found in a two-component spin-orbit coupled BEC
subject to an in-plane gradient magnetic field \cite{Zhou2}. However, the
present toroidal system can exhibit more rich and complex topological
structures due to the interplay among rotation, interatomic interactions,
SOC, and toroidal confinement.

In addition, a topology transition is indicated in the system. We take the
case of $\beta _{12}>\beta _{11}$ as an example. In figure 3(c), there exist
two vortices around the center of component $1$. For comparison, there are
five visible vortices around the center visible vortex of component $1$ in
figure 3(d) and eight nearest hidden vortices around the center giant vortex
of component $1$ in figure 3(e), respectively. Therefore these patterns
display the transition of the topological structure as the strength of SOC
increases, which indicates that SOC can be used to control the topological
structure of the rotating toroidal BEC. Note that there are obvious singular
points in the total density $|\psi _{1}|^{2}+|\psi _{2}|^{2}$ of the system
(see the fifth column). Thus the topological defects in the present system
are different from the common coreless vortices \cite{Matthews} or the
so-called Anderson-Toulouse vortices \cite{Anderson}.

The corresponding spin-density distributions are shown in figure 4. In the
pseudo-spin representation, the blue region denotes spin-down and the red
region denotes spin-up. To understand the interesting spin structures, we
take the case of $\Omega =0.9$, $\beta _{12}=50$, and $\lambda _{x}=\lambda
_{y}=1$ as a typical example. From $S_{z}$ in figure 4(a), the spin
continuously turn from down to up along the radial direction and then
gradually turn back to down. At the same time, $S_{x}$ obeys an odd parity
distribution along the $x$ direction and an even parity distribution along
the $y$ direction, but the situation is the reverse for $S_{y}$.

\begin{figure}[tbh]
\centerline{\includegraphics*[width=12cm]{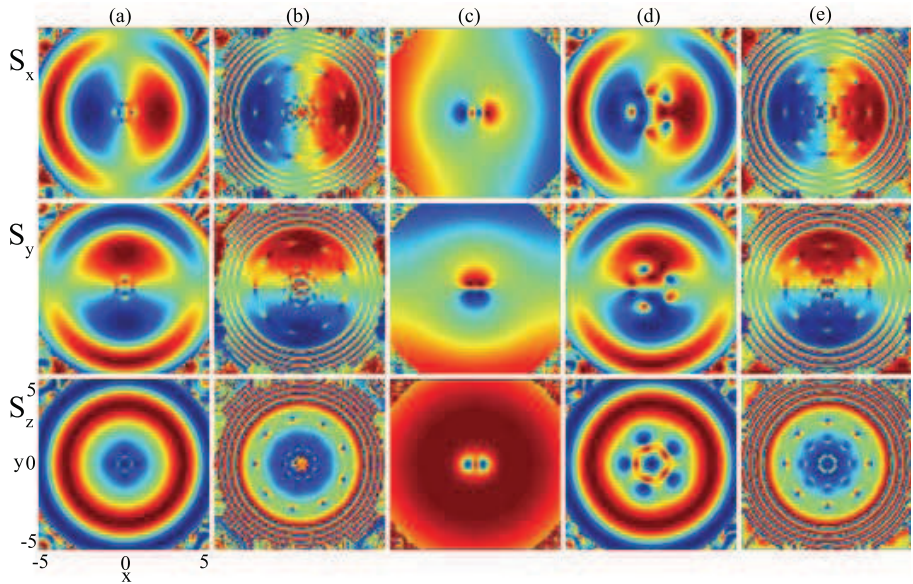}}
\caption{(Color online) Spin densities of the rotating interacting spin-$1/2$
BECs with SOC in a toroidal trap, where $\Omega =0.9$, and columns (a)-(e)
correspond to rows (a)-(e) in figure 3, respectively. The rows (from top to
bottom) denote $S_{x}$, $S_{y}$ and $S_{z}$ components of the spin density
vector, respectively. The horizontal and vertical coordinates $x$ and $y$
are in units of $a_{0}$.}
\label{Figure4}
\end{figure}

Displayed in figure 5(a) and figures 5(b)-5(c) are the topological charge
density and the spin texture for the parameters in figure 4(a),
respectively. Obviously, there are five radial-out skyrmions \cite%
{Liu,Skyrme,Zhai2,Kasamatsu3} in figure 5(b) which constitute a skyrmion
lattice in spin representation. The larger SOC leads to the generation of a
more complex skyrmion lattice (including more skyrmions even giant skyrmion
\cite{Yang}) through the strong interaction between the `spin' angular
momentum and orbit angular momentum of the atoms, where the necklace-like
distribution in the spin density becomes more distinct. This characteristic
is present evidently for both the cases of $\beta _{12}<\beta _{11}$ and $%
\beta _{12}>\beta _{11}$ [see figures 4(b), 4(d) and 4(e)]. In particular,
we find that at certain parameters the rotating toroidal system with SOC
supports a skyrmion pair. Figures 5(e)-5(f) show the spin texture and its
local amplification at $\Omega =0.9$, $\beta _{12}=150$, and $\lambda
_{x}=\lambda _{y}=0.2$, and figure 5(d) gives the corresponding topological
charge density. We see that the spin texture exhibits a
radial(out)-radial(out) skyrmion configuration, which is quite different
from the normal seven basic skyrmion configurations \cite{Liu} of rotating
spin-orbit-coupled BECs in a harmonic trap. Our numerical calculation shows
that the topological charge approaches $\left\vert Q\right\vert =2$, and
thus we may call it skyrmion pair.

\begin{figure}[tbh]
\centerline{\includegraphics*[width=10cm]{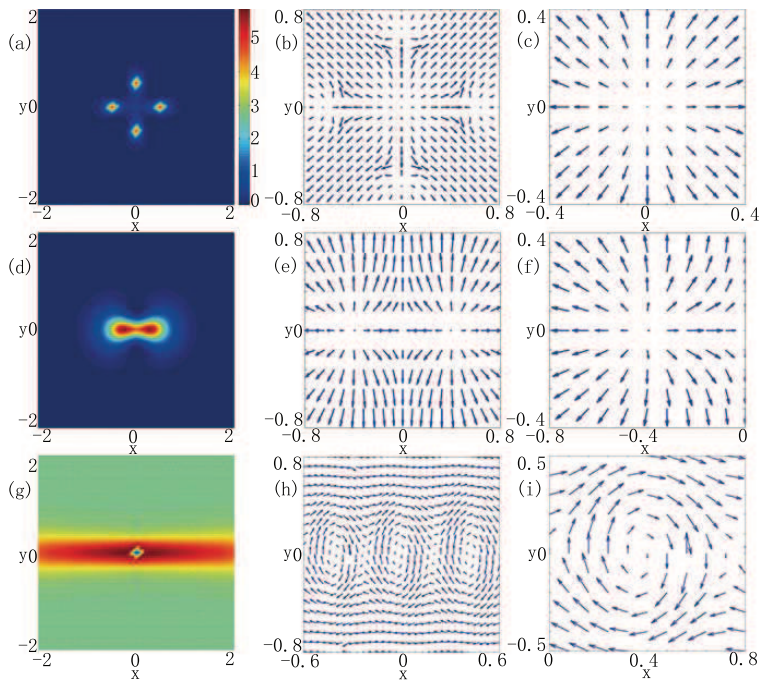}}
\caption{(Color online) Topological charge densities and spin textures of
rotating spin-$1/2$ BEC with SOC in a toroidal trap. The left column denotes
topological charge density, the middle column is the corresponding spin
texture, and the right column is the local amplification of the spin
texture. (a)-(c) $\Omega =0.9$, $\protect\beta _{12}=50$, $\protect\lambda %
_{x}=\protect\lambda _{y}=1$, (d)-(f) $\Omega =0.9$, $\protect\beta %
_{12}=150 $, $\protect\lambda _{x}=\protect\lambda _{y}=0.2$, and (g)-(i) $%
\Omega =1.2$, $\protect\beta _{12}=150$, $\protect\lambda _{x}=2$, $\protect%
\lambda _{y}=0$. The horizontal and vertical coordinates $x$ and $y$ are in
units of $a_{0}$.}
\label{Figure6}
\end{figure}

\subsubsection{Fixed intercomponent interaction}

Now we discuss the ground state of the system at fixed intercomponent
interaction. Figure 6 and figure 7 depict the density distributions and
phase distributions obtained under various strengths of SOC and rotation
frequencies for the initially miscible BECs ($\beta _{12}=50$) and the
initially immiscible BECs ($\beta _{12}=150$), respectively. In the
initially miscible case of $\beta _{12}=50$, when the SOC strength is
relatively weak and the rotation frequency is small, there is a typical
triangular vortex lattice in component $1$ and a doubly quantized vortex in
component $2$. With the increase of rotation frequency, the trap center
region is pinned by a triply quantized vortex for component $1$ and a doubly
quantized vortex for component $2$. At the same time, a vortex ring is
formed on the periphery of the giant vortex in each component. For large SOC
strength, each component forms a complex topological structure composed of
laminar visible vortex ring(s) and a large density hole. Note that the large
density hole represents a giant vortex plus hidden vortex ring(s) rather
than a pure giant vortex which is observed in conventional rotating BECs
\cite{Fetter2}. In addition, the density distributions and phase
distributions for the two components become similar. Therefore the
topological structure of the system is strongly influenced by the interplay
among the SOC, rotation frequency, the interparticle interactions, and the
toroidal confinement.

\begin{figure}[tbh]
\centerline{\includegraphics*[width=8cm]{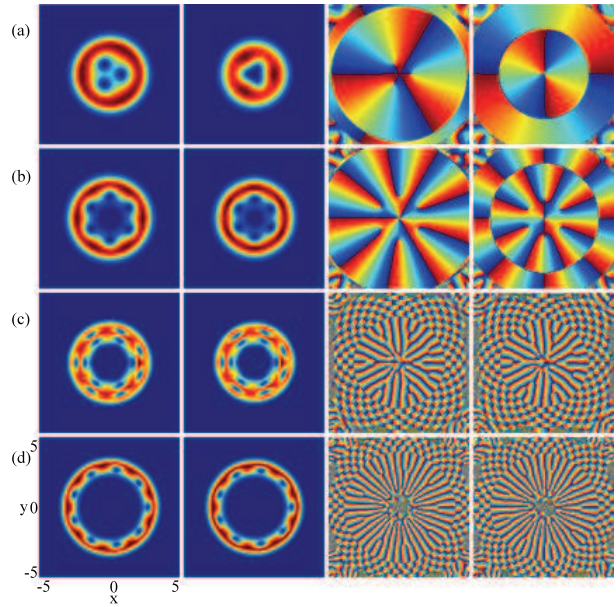}} 
\caption{(Color online) Ground states of rotating interacting spin-$1/2$
BECs with SOC in a toroidal trap, where $\protect\beta _{12}=50$. (a) $%
\protect\lambda _{x}=\protect\lambda _{y}=1$, $\Omega =0.4$, (b) $\protect%
\lambda _{x}=\protect\lambda _{y}=1$, $\Omega =1.2$, (c) $\protect\lambda %
_{x}=\protect\lambda _{y}=10$, $\Omega =0.4$, and (d) $\protect\lambda _{x}=%
\protect\lambda _{y}=10$, $\Omega =1.2$. The columns (from left to right)
represent $|\protect\psi _{1}|^{2}$, $|\protect\psi _{2}|^{2}$, arg$\protect%
\psi _{1}$, and arg$\protect\psi _{2}$, respectively. The horizontal and
vertical coordinates $x$ and $y$ are in units of $a_{0}$.}
\label{Figure5}
\end{figure}

In the initially immiscible case of $\beta _{12}=150$, when the SOC is
relatively weak and the rotation frequency is small, the two components
exhibit remarkable phase separation as shown in figure 7(a). The topological
structure of the system in figure 7(a) is a typical Anderson-Toulouse
coreless vortex \cite{Anderson}, where the core of the circulating external
component is filled with the other nonrotating component. With the further
increase of rotation frequency, each component of the system evolves from a
standard triangular vortex lattice into a special topological configuration
which is comprised of an exterior vortex ring and a central vortex (or giant
vortex) due to the presence of central barrier [see figures 7(b) and 7(c)].
For large SOC strength and small rotation frequency, the visible vortices
tend to be elongated along the radius and linked one after another [figure
7(d)]. When the rotation frequency increases, the elongated effect of the
visible vortices becomes deformed and eventually disappears, and the final
topological structure of the system is similar to that in the initially
miscible case. From figure 2 to figure 7, we show that the rich topological
structures of the rotating toroidal BECs with SOC are determined by the
interplay among the SOC, the rotation frequency, and the interatomic
interactions, especially for the SOC strength.

\begin{figure}[tbh]
\centerline{\includegraphics*[width=11cm]{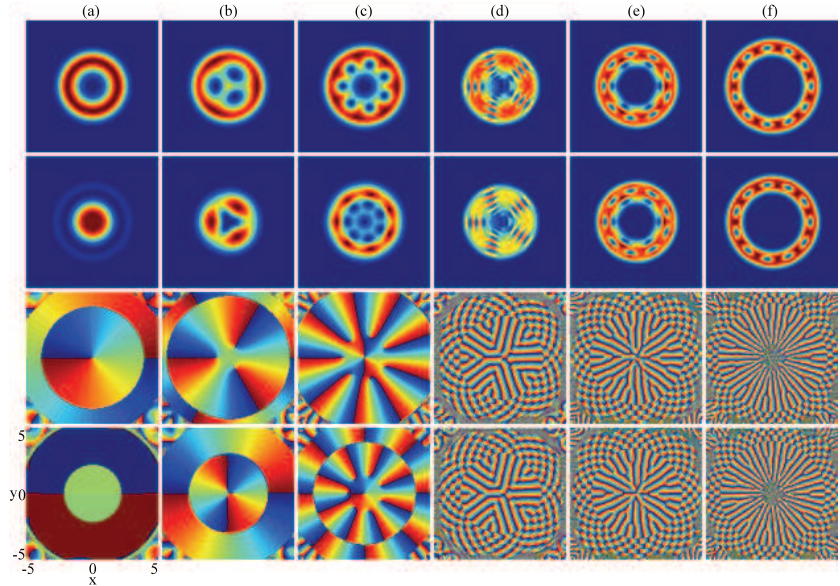}}
\caption{(Color online) Ground states of rotating interacting spin-$1/2$
BECs with SOC in a toroidal trap, where $\protect\beta _{12}=150$. (a) $%
\protect\lambda _{x}=\protect\lambda _{y}=1$, $\Omega =0.2$, (b) $\protect%
\lambda _{x}=\protect\lambda _{y}=1$, $\Omega =0.6$, (c) $\protect\lambda %
_{x}=\protect\lambda _{y}=1$, $\Omega =1.2$, (d) $\protect\lambda _{x}=%
\protect\lambda _{y}=10$, $\Omega =0.2$, (e) $\protect\lambda _{x}=\protect%
\lambda _{y}=10$, $\Omega =0.6$, and (f) $\protect\lambda _{x}=\protect%
\lambda _{y}=10$, $\Omega =1.2$. The rows (from top to bottom) represent $|%
\protect\psi _{1}|^{2}$, $|\protect\psi _{2}|^{2}$, arg$\protect\psi _{1}$,
and arg$\protect\psi _{2}$, respectively. The horizontal and vertical
coordinates $x$ and $y$ are in units of $a_{0}$.}
\label{Figure7}
\end{figure}

\subsubsection{One-dimensional SOC}

Finally, we study the ground-state structures of rotating toroidal BECs in
the presence of 1D SOC, where the relevant parameters are $\beta _{12}=150$,
$\lambda _{x}=2$, $\lambda _{y}=0$, $\Omega =0.5$ [figure 8(a)] and $\Omega
=1.2$ [figure 8(b)]. The columns (from left to right) in figures 8(a)-8(b)
represent $|\psi _{1}|^{2}$, $|\psi _{2}|^{2}$ , arg$\psi _{1}$, arg$\psi
_{2}$, $|\psi _{1}|^{2}+|\psi _{2}|^{2}$, and $|\psi _{1}|^{2}-|\psi
_{2}|^{2}$, respectively. The corresponding spin density components $S_{x}$,
$S_{y}$ and $S_{z}$ are shown in the left three columns and the right three
columns in figure 8(c), respectively. When $\Omega =0.5$, there is an
obvious visible vortex string along the $y=0$ axis in each component due to
the 1D SOC along the $x$ direction, and the system displays an evident phase
separation [see figure 8(a)]. As shown in figure 8(b), besides the vortex
string along the $y=0$ axis, there exists a doubly-quantized vortex in the
trap center and a vortex distribution along the $x=0$ axis in individual
components. The main reason is that for large rotation frequency $\Omega
=1.2 $ the $x$-direction vortex string resulted from the combined effect of
1D SOC and rotation can only carry partial angular momentum and the remain
angular momentum is inevitably carried by the central giant vortex and the
transverse vortices beside the $y=0$ axis.

\begin{figure}[tbh]
\centerline{\includegraphics*[width=12cm]{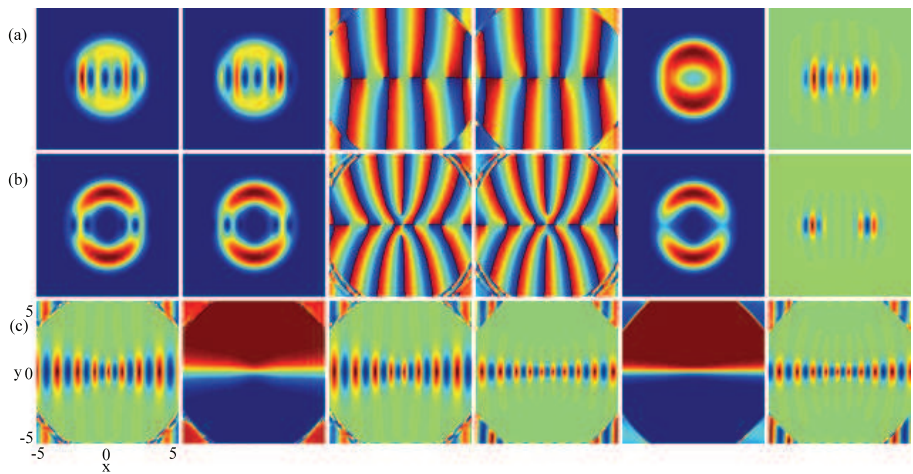}}
\caption{(Color online) Ground states and spin densities of rotating
interacting BECs with 1D SOC. The six columns (from left to right) in rows
(a)-(b) represent $|\protect\psi _{1}|^{2}$, $|\protect\psi _{2}|^{2}$ , arg$%
\protect\psi _{1}$, arg$\protect\psi _{2}$, $|\protect\psi _{1}|^{2}+|%
\protect\psi _{2}|^{2}$, and $|\protect\psi _{1}|^{2}-|\protect\psi %
_{2}|^{2} $, respectively. The left three and right three columns in (c)
denote $S_{x}$, $S_{y}$ and $S_{z}$ components of spin density vectors
corresponding to the first and the second rows, respectively. Here the
parameters are $\protect\beta _{12}=150$, $\protect\lambda _{x}=2$, and $%
\protect\lambda _{y}=0$. The rotation frequencies in (a) and (b) are $\Omega
=0.5$ and $\Omega =1.2$, respectively. The horizontal and vertical
coordinates $x$ and $y$ are in units of $a_{0}$.}
\label{Figure9}
\end{figure}

In the spin representation, the spin component $S_{y}$ develops into two
remarkable spin domains due to the phase separation of the two component
densities, and the boundary between the two spin domains forms a spin domain
wall with $\left\vert S_{y}\right\vert \neq 1$, which can be seen clearly in
figure 8(c). It is well known that the spin domain wall for a nonrotating
two-component condensate system is a typically classical Neel wall, where
the spin flips only along the vertical direction of the wall. However, our
numerical simulation\ of the spin texture shows that in the region of spin
domain wall the spin flips not only along the $y$ direction (the vertical
direction of domain wall) but also along the $x$ direction (the domain-wall
direction), which implies that the observed spin domain wall is a new type
of domain wall. The topological charge density and the spin texture for the
case of $\beta _{12}=150$, $\lambda _{x}=2$, $\lambda _{y}=0$ and $\Omega
=1.2$ are shown in figures 5(g)-5(i), where a special topological structure
of skyrmion string is formed in the spin representation. In addition, the $%
S_{x}$ and $S_{z}$ components of the spin density display a chain structure,
where $S_{x}$ satisfies the even-parity distribution while $S_{z}$ obeys the
odd-parity distribution.

\section{Conclusion}

In summary, we investigate systematically the ground state of rotating
two-component BECs with Rashba spin-orbit coupling in a toroidal trap. The
influence of rotation, SOC, and interatomic interactions on the ground state
of the system is analyzed in detail. In the absence of SOC, large rotation
velocity yields the formation of giant vortices for initially miscible
two-component BECs. However, for initially immiscible two-component BECs
rapid rotation can generate semiring density structures with irregular
hidden vortices. In the absence of rotation, large strength of 2D isotropic
SOC may result in a heliciform-stripe phase for initially separated
two-component BECs. The combined effects of SOC, rotation, and interparticle
interactions are discussed. For large rotation frequency, strong 2D SOC
favors a ringlike visible vortex configuration, where the region of large
density hole is occupied by a hidden giant vortex and a (several) hidden
vortex ring(s). In particular, strong 1D anisotropic SOC leads to the
generation of hidden linear vortex string and the transition between phase
separation (phase mixing) and phase mixing (phase separation). In addition,
complex spin topological structures, such as skyrmion pair, skyrmion string,
and skyrmion lattice are found in the present system. This work provides
exciting perspectives for topological excitations in quantum gases and
condensed matter physics.

\textit{Note added:} Recently, we became aware of two preprints by Zhang
\textit{et al.} \cite{XFZhang} and White \textit{et al.} \cite{White}, which
have also studied some relevant properties of spin-orbit coupled BECs in a
toroidal trap.

\begin{acknowledgments}
L.W. thanks Chuanwei Zhang, Hui Zhai, Yongping Zhang, Zhi-Fang Xu, and
Xiang-Fa Zhou for helpful discussions. This work is supported by the
National Natural Science Foundation of China (Grant No. 11475144), the
Natural Science Foundation of Hebei Province of China (Grant No.
A2015203037), and Ph.D. foundation of Yanshan University (Grant No. B846).
\end{acknowledgments}

\end{document}